\documentclass[a4paper,11pt]{article}
\usepackage{aas_macros}
\usepackage{physics}
\usepackage{jcappub}
\usepackage{pgfplots}
\pgfplotsset{compat=1.18}

\usepackage{amsmath,amssymb, amsthm,amstext,comment,lipsum,mathtools,placeins}
\usepackage{natbib}
\usepackage{graphicx}
\usepackage{xcolor}
\usepackage{array, enumerate}
\usepackage{bm}
\usepackage{multirow}
\usepackage{tikz}
\usetikzlibrary{arrows.meta, decorations.markings}

\def\apjl{Astroph.J.Lett.}
\def\apjs{Astroph.J.S.}

\def\aap{Astronomy\&Astrophysics}

\def\bfx{\mathbf{x}}

\def\bfthe{\boldsymbol{\theta}}

\def\bfb{\mathbf{b}}
\def\rmd{{\rm d}}

\def\bfd{\mathbf{d}}
\def\bfC{\mathbf{C}}

\def\bfD{\mathbf{D}}

\def\bfM{\mathbf{M}}
\def\bfN{\mathbf{N}}

\def\bftheta{\boldsymbol{\theta}}

\newcommand{\VEV}[1]{\left\langle#1\right\rangle}

\newcommand{\refEq}[1]{Eq.~(\ref{eq:#1})}
\newcommand{\refeqs}[2]{Eqs.~(\ref{eq:#1})--(\ref{eq:#2})}          
\newcommand{\reffig}[1]{Fig.~\ref{fig:#1}}          
\newcommand{\refsec}[1]{Sec.~\ref{sec:#1}}
\newcommand{\refapp}[1]{App.~\ref{app:#1}}

\begin{document}

\title{How to augment cosmic shear measurements with radio polarimetry of galaxies?}

\author[a]{Liang Dai,}
\author[b]{Junwu Huang,}
\author[a]{Weichen Winston Yin,}
\author[e]{Rui Zhou,}
\author[c,a]{and Simone Ferraro}

\affiliation[a]{Department of Physics, University of California, 366 Physics North MC 7300, Berkeley, CA. 94720, USA}
\affiliation[b]{Perimeter Institute for Theoretical Physics, 31 Caroline St.~N., Waterloo, Ontario N2L 2Y5, Canada}
\affiliation[c]{Physics Division, Lawrence Berkeley National Laboratory, Berkeley, CA 94720, USA}
\affiliation[d]{Berkeley Center for Cosmological Physics, Department of Physics, University of California, Berkeley, CA 94720, USA}
\affiliation[e]{School of Physics, University of Melbourne, Parkville, VIC, 3010, Australia}

\emailAdd{liangdai@berkeley.edu}
\emailAdd{jhuang@perimeterinstitute.ca}
\emailAdd{rui.zhou@student.unimelb.edu.au}

\abstract{
The integral polarization of spiral galaxies in the radio band has been proposed as a new tracer of the intrinsic galaxy shape that augments lensing shear measurements. We revisit the method of shear estimation in this context. We introduce a new statistical model in which galaxy shape and polarization are Gaussian random variables with their covariance characterizing the quality of polarization-shape alignment. Applying the principle of likelihood maximization, we then analytically derive unbiased, minimal-variance estimators, which allow to simultaneously estimate gravitational shear, intrinsic shape alignment and line-of-sight polarization rotation, all at once and accurate to first order in these three effects. New to the literature, our estimators have the merits of being free of biases, robust in situations of few galaxies or poor polarization-shape alignment, allowing analytic reconstruction noise covariance, and minimizing uncertainties in power spectrum estimation, thus resolving conceptual issues of the existing estimation methods. This new analytic framework is generally applicable to future research that exploits the polarization-shape alignment effect of galaxies.

}

\maketitle

\section{Introduction} 
\label{sec:intro}

In star-forming galaxies, the integral ISM continuum emission in the radio band ($\nu \sim 1$--$10\,$GHz; often dominated by synchrotron emission except toward high frequencies) has a polarization direction that strongly correlates with the apparent minor axis of the galaxy in optical images. Stil {\it et al} first empirically discovered this correlation at $4.8\,$GHz in small samples of nearby galaxies~\citep{stil2009integrated}. This correlation arises because the interstellar magnetic field that locally sets the polarization direction of the diffuse synchrotron emission has an ordered global component tracing the geometry of the gas disk~\citep{stil2009integrated, Sun2012PolMWlikeGalaxies}. Therefore, this effect should be prevalent in distant galaxies. Indeed, Zhou {\it et al} confirmed and quantified this effect for simulated galaxies in the \texttt{IllustrisTNG50} project over a range of redshifts ($z=0$--$2$) and for a range of observed radio frequencies ($\nu=1$--$8\,$GHz)~\citep{Zhou2025TNG50polshape}. Despite several public datasets from radio continuum surveys with polarimetry~\citep{Lacy:2019rfe, Taylor2024MIGHTEEfields}, sufficiently deep radio data for a cosmologically significant sample of galaxies are still lacking, which would be needed for accurately modeling the statistics of polarization-shape (mis)alignment.

Recently, there has been renewed community attention to cosmology and fundamental physics applications that exploit polarization-shape alignment in galaxies, with the prospect of large datasets to be delivered by upcoming radio-band galaxy surveys such as those from the Square Kilometre Array (SKA) project~\citep{Jarvis2015CosmologyWithSKA}. When leveraging radio continuum imaging of distant galaxies, the integral polarization is measurable without the need to spatially resolve individual galaxies. At the same time, galaxy shape measurements are readily available from current and upcoming optical imaging surveys since shapes are standard inputs to weak lensing analyses. 

Polarization-shape alignment for general extragalactic sources has been found useful in the search of cosmic birefringence. Here, the observed source shape is used as a tracer of the intrinsic polarization direction~\citep{Kamionkowski2010CosmicBirefringenceAGN, Whittaker2017measuring, yin2025new}. A synergy of polarimetric radio surveys and optical shape surveys will enable us to detect or constrain polarization rotation along the line of sight induced by ultralight axion fields in the Universe~\citep{Carroll:1989vb,Carroll1998PhRvL..81.3067C, Lue1999PhRvL..83.1506L, Arvanitaki2010PhRvD..81l3530A, Agrawal:2019lkr, Namikawa2025PhRvL.135p1004N}, at sensitivities that rival Stage-IV CMB experiments~\citep{yin2025new} and enabling redshift tomography~\citep{Naokawa2026CosmicBirefringenceTomography}.

For another application, it has been recognized that  a novel method to measure weak gravitational lensing of background galaxies can be developed based on polarization-shape alignment~\citep{BrownBattye2011weaklensingIA, BrownBattye2011WeakLensingMapping, Jarvis2016MIGHTEEsurvey, Harrison2016SKAweaklensingI}. Compared to measuring lensing shear simply from the average observed galaxy shape, polarization supplements information on the intrinsic galaxy shape, which not only reduces the shape noise but also tells apart gravitational shear and intrinsic shape alignment~\citep{Croft2000IntrinsicAlignment,Heavens2000IntrinsicAlignment,Catelan2001IntrinsicAlignment, Crittenden2001IntrinsicAlignment, Jing2002HaloIntrinsicAligment, Mackey2002IAtheory}. This conceptually parallels other promising ideas on augmenting shear measurements using additional galaxy observables, such as the method of kinematic weak lensing~\citep{Blain2002CosmicShearWithSpectralImaging, Morales2006WeakLensingWithVelocityMaps, Xu2023KinematicWeakLensingRST}. More than a cosmological nuisance, intrinsic alignment may carry unique information about galaxy formation~\citep{Chisari2025AAPRreview}, primordial non-Gaussianity~\citep{Chisari2013IAInflationaryPNG,Schmidt2015InflationImprintsIA, Taruya2020IAConstraintsOnCosmology,Akitsu2021PNGsim,Kurita2026ParityViolatingPNGinIA}, or primordial tensor modes~\citep{Schmidt2012CosmicRulers, Dai2013pGWquadPk, Schmidt2014pGWtides, Philcox2024NewPhysicsInGalaxyShapes}, so a new independent capability to measure it will be valuable. 

In the pioneer work of Ref.~\citep{BrownBattye2011weaklensingIA}, Brown \& Battye (BB11) developed new shear estimators that leverage radio polarization information. In constructing these estimators, the authors use the integral polarization direction as an indicator of the intrinsic galaxy minor axis. Aiming to remove degeneracy between lensing shear and intrinsic alignment, they choose to project out the ellipticity component aligned with that direction. This reduces the shape noise, yet causes intrinsic alignment to contaminate shear estimation by a small amount that precisely vanishes only in the limit of perfect polarization-shape alignment (i.e. when the polarization-shape misalignment angle, defined in \refEq{deltheta_def}, has zero standard deviation). While this bias may appear numerically insignificant when intrinsic alignment is weak for the relevant galaxy sample, it poses a conceptual problem, which can become important with large statistics from future galaxy data. 

Later in Ref.~\citep{Whittaker2015SeparateShearIA}, Whittaker, Brown \& Battye (WBB15) identified the root of this bias to be imperfect knowledge of the intrinsic shape due to randomness in the polarization direction, and derived an analytic approximation for the bias. This then enabled the authors to modify the original BB11 estimators in \citep{BrownBattye2011weaklensingIA} so that shear bias is removed at the level of combining measurements done for many independent galaxies. Ref.~\citep{BrownBattye2011WeakLensingMapping} demonstrated that the spatially varying weak lensing convergence can be correctly mapped out using these estimators. Ref.~\cite{Whittaker2017measuring} further generalized the method to estimating an additional polarization rotation effect along the line of sight, but focused on applying the estimators to resolved AGN radio lobes.
However, corrections are implemented in the WBB15 estimators at the cost of inflated reconstruction noise compared to the BB11 method, due to estimation outliers that result from inverting (near-)singular matrices. This occurs when the number of usable galaxies in each pixel of map-making is small~\citep{Whittaker2015SeparateShearIA}, an issue absent in the original BB11 estimators. This in principle poses a limitation for reconstructing shear modes on very small angular scales or in shallow radio continuum surveys~\citep{Whittaker2015SeparateShearIA}.

Following an independent logic, we revisit the problem of separating shear and intrinsic alignment with polarization information and that of simultaneously estimating extrinsic polarization rotation. Building upon but significantly revising the mathematical framework introduced in Ref.~\citep{yin2025new}, we first present an analytic model describing galaxy ellipticity and integral polarization as Gaussian random variables. Crucially, polarization-shape alignment is captured by a covariance between ellipticity and polarization. For simplicity, we will not further model the strong dependence of polarization-shape alignment on galaxy inclination~\citep{stil2009integrated}, only to note that this dependence is theoretically expected~\citep{stil2009integrated} and confirmed for simulated galaxies~\citep{Zhou2025TNG50polshape}.

We will then analytically derive unbiased, minimal-variance estimators, for shear, intrinsic alignment, and polarization rotation, accurate to first order in these three effects and all at once. Our derivation is guided by likelihood maximization, which is a powerful first principle already proven successful in solving important inference problems such as CMB weak lensing reconstruction~\citep{Hirata_2003weaklensing, Hirata_2003} and cosmic birefringence detection~\citep{Yin:2021kmx}. Despite their remarkably simple forms, the new estimators we obtain are different from those found in the literature. They are unbiased, robustly applicable with any number of galaxies, and have optimized reconstruction noises. For a proof of concept, we will validate them using toy-model galaxy mocks. A key result of this work will be an improved and generalized analytical framework for statistical estimation than presented in Ref.~\citep{yin2025new}, in light of improved qualitative and quantitative understanding of the polarization-shape alignment effect~\citep{Zhou2025TNG50polshape}. Analogous to reconstructing the weak lensing convergence field from primary CMB temperature and polarization anisotropies, these estimators can be used to reconstruct sky maps of lensing shear, intrinsic alignment and polarization rotation, from which angular power spectra can be measured.

The remainder of this paper is organized as follows. In \refsec{theory}, we introduce zero-mean, correlated Gaussian distributions to model the intrinsic statistics of galaxy ellipticity and polarization. Then in \refsec{shear_IA}, we will model how lensing shear and intrinsic alignment, which we treat as effects extrinsic to galaxy internal structure, bias the Gaussian distributions of shape and polarization. The corresponding analytic expressions for the likelihood functions are written down in \refsec{likelihood}, with or without being modified by extrinsic effects. In the following \refsec{optest}, we analytically derive a set of unbiased, minimal-variance estimators for lensing shear, polarization rotation, and intrinsic alignment, along with their noise covariance under the null hypothesis, which form the key results of this work. In \refsec{noise}, our analytic results are generalized to account for measurement errors in galaxy shape and polarization.
In \refsec{mocks}, we validate our estimators by applying them to mock samples of galaxy shape and polarization, and demonstrate, as a proof of concept, how the angular power spectra of shear, intrinsic alignment and polarization rotation can be best measured with our new estimators. We will give concluding remarks in \refsec{concl}. For readers interested in calculational details, we outline in \refapp{est_from_likelihood} a derivation of the minimal-variance quadratic estimators from the principle of maximal likelihood estimation. In \refapp{est_power_spectra}, we discuss an algorithm to estimate auto- and cross-angular power spectra and the associated statistical uncertainties involving multiple correlated Gaussian random quantities on the sky. This algorithm is applied in our mock tests.

\section{Intrinsic statistics of galaxy shape and polarization} 
\label{sec:theory}

With the goal to quantitatively study how shear, intrinsic alignment and polarization rotation can be simultaneously and optimally estimated, we develop a novel mathematical framework. In this section, we first introduce a new statistical model describing random galaxy shapes and polarizations, parameterizing imperfect polarization-shape alignment. Then, we will derive estimators for shear, intrinsic alignment, and rotation.

Consider a survey in which two spin-2 properties are measured for every galaxy: ellipticity components $(\varepsilon_1,\,\varepsilon_2)$ describing the galaxy shape, and Stokes parameters $(q,\,u)$ describing the integral polarization. Ellipticity variables $(\varepsilon_1,\,\varepsilon_2)$ are typically measured from optical images of galaxies, but in principle can also be measured in radio imaging surveys if the radio emission of the galaxy is spatially resolved. The Stokes parameters $(q,\,u)$ further require radio polarimetry. Since star-forming galaxies have vastly different radio continuum luminosities~\citep{Hansen2024SHARKmodel}, it is preferable to normalize them by the integral intensity $I$, i.e. we define $q=Q/I$ and $u=U/I$, in order to render the distributions of $q$ and $u$ more Gaussian~\citep{yin2025new}.

Our basic assumption is that for the galaxy sample $(\varepsilon_1, \,\varepsilon_2,\,q,\,u)$ can be approximated as Gaussian random variables. This assumption is different from Ref.~\citep{yin2025new}, where it is assumed instead that the parity-even and parity-odd products of ellipticity and polarization have Gaussian distributions. An analysis of the \texttt{IllustrisTNG} simulated galaxies suggests that those products have non-Gaussian distributions~\citep{Zhou2025TNG50polshape}.

In the absence of lensing shear or intrinsic alignment, galaxy ellipticity and polarization have no preferred orientations on the sky. In this case, the spin-2 variables $(\varepsilon_1, \,\varepsilon_2,\,q,\,u)$ all have zero means, i.e. $\VEV{\varepsilon_1} = \VEV{\varepsilon_2} = 0$ and $\VEV{q}=\VEV{u}=0$~\footnote{The notation $\VEV{\cdots}$ stands for ensemble average of many galaxies.}.

Ellipticity and polarization variables can be reparametrized using complex numbers
\begin{align}
    q + i\,u & = p\,e^{2\,i\,\theta_p}, \\
    \varepsilon_1 + i\,\varepsilon_2 & = \varepsilon\,e^{2\,i\,\theta_e}.
\end{align}
Here $p=\sqrt{q^2+u^2}$ is the polarization degree, $\varepsilon = \sqrt{\varepsilon^2_1 + \varepsilon^2_2}$ is magnitude of ellipticity, and $\theta_p$ and $\theta_e$ are the position angles of the integral polarization and that of the shape ellipse on the sky, respectively. The polarization-shape misalignment angle is defined as~\cite{yin2025new}
\begin{align}
\label{eq:deltheta_def}
    \delta\theta = \theta_p - \theta_e - \frac{\pi}{2}, \quad -\frac{\pi}{2} \leqslant \delta\theta \leqslant \frac{\pi}{2}.
\end{align}
Thus, $\delta\theta=0$ corresponds to the net linear polarization direction aligned with the minor axis of the shape ellipse.

The nonzero covariances of the shape and polarization observables $(\varepsilon_1, \,\varepsilon_2,\,q,\,u)$ are parameterized in general as the following:
\begin{align}
\label{eq:qu_var}
\VEV{q^2} = \VEV{u^2} = & \sigma^2_p, \\
\label{eq:ellip_var}
    \VEV{\varepsilon^2_1} = \VEV{\varepsilon^2_2} = & \sigma^2_\varepsilon, \\
\label{eq:qu_ellip_covar}
    \VEV{q\,\varepsilon_1} = \VEV{u\,\varepsilon_2} = & - \sigma_p\,\sigma_\varepsilon\,\cos\delta,
\end{align}
where $\sigma_p$ and $\sigma_\varepsilon$ are standard deviations for the ellipticity and polarization, respectively. Covariance entries $\VEV{q\,u}$ and $\VEV{\varepsilon_1\,\varepsilon_2}$ are forbidden by statistical isotropy. Other covariance entries like $\VEV{q\,\varepsilon_2}$ and $\VEV{u\,\varepsilon_1}$ must also vanish. These would change sign under spatial reflection, but the distributions of galaxy properties are not known to violate parity. Thus, \refeqs{qu_var}{qu_ellip_covar} describe the most general {\it intrinsic} correlations for individual galaxies consistent with fundamental symmetries.

In particular, \refEq{qu_ellip_covar} reflects the key feature that polarization and shape have correlated directions~\citep{stil2009integrated,Zhou2025TNG50polshape}. The degree of this correlation is parametrized by $\delta$ ($0 \leqslant \delta \leqslant \pi/2$). Integral polarization aligns perfectly with the minor (hence the minus sign in \refEq{qu_ellip_covar}) axis of the shape ellipse if $\delta=0$. If $\delta=\pi/2$, the two directions are unrelated.

Under the assumption of Gaussian statistics, the intrinsic shape and polarization variables are drawn from a multi-variate, unbiased Gaussian distribution:
\begin{align}
\label{eq:fullPDF_noIA}
& P(q,\,u,\,\varepsilon_1,\,\varepsilon_2)\,\rmd q\,\rmd u\,\rmd\varepsilon_1\,\rmd\varepsilon_2 = \nonumber\\
& \frac{4}{(2\pi)^2\,\sigma^2_p\,\sigma^2_\varepsilon\,\sin^2\delta} \times \exp\left[-\frac{\varepsilon^2}{2\,\sigma^2_\varepsilon\,\sin^2\delta} - \frac{p^2}{2\,\sigma^2_p\,\sin^2\delta} + \frac{\varepsilon\,p\,\cos\delta}{\sigma_p\,\sigma_\varepsilon\,\sin^2\delta}\,\cos2\delta\theta\right]\,\,p\,\rmd p\,\varepsilon\,\rmd\varepsilon\,\rmd\delta\theta\,\rmd\theta_e,
\end{align}
where we have transformed the variables to an equivalent set $(p,\,\varepsilon,\,\delta\theta,\,\theta_e)$.
Without a preferred galaxy orientation on the sky, $\theta_e$ has a uniform distribution. Integrating out $p$, $\varepsilon$ and $\theta_e$,  we derive an analytic PDF for the misalignment angle,
\begin{align}
\label{eq:Delta_theta_pdf_model}
    P(\delta\theta;\,\delta) = & \frac{\sin^2\delta}{\pi}\,\left[\frac{\left(\pi-\arccos(\cos (2\delta\theta)\,\cos\delta) \right)\,\cos (2\delta\theta)\,\cos\delta}{\left(1-\cos^2 (2\delta\theta)\,\cos^2\delta\right)^{3/2}} + \frac{1}{1-\cos^2(2\delta\theta) \,\cos^2\delta}\right],
\end{align}
which is set by a single parameter $\delta$ and is normalized for $-\pi/2 \leqslant \delta\theta \leqslant \pi/2$. Ref.~\cite{Zhou2025TNG50polshape} used the analytic distribution \refEq{Delta_theta_pdf_model} to fit the statistics of polarization-shape (mis-)alignment calculated for simulated galaxies in the \texttt{IllustricTNG50} data release~\citep{nelson2021illustristngsimulationspublicdata}. The analytic model fits the misalignment angle distribution well for a wide range of galaxy redshifts, disk inclinations, and observed radio frequencies. Here, we shed light on the derivation of this analytic distribution to fill this literature gap.

Ref.~\citep{Zhou2025TNG50polshape} showed that galaxies viewed at different inclination angles are fit by significantly different $\delta$ values. The $\delta$ value decreases for more inclined galaxies, a behavior that is theoretically anticipated~\citep{stil2009integrated}. In this work, however, we do not attempt to construct more sophisticated statistical models of galaxy shape and polarization to capture this correlation. We shall adopt the simplistic assumption that a single $\delta$ value applies to the entire galaxy sample, while generalizing our framework to multiple sub-samples with different $\delta$ values will be straightforward. In practice, the value of $\delta$ can be empirically determined for a general galaxy (sub-)sample by fitting the observed distribution of misalignment angles.

\section{Effects of shear and intrinsic alignment} 
\label{sec:shear_IA}

The spin-2 variables $(\varepsilon_1, \,\varepsilon_2,\,q,\,u)$ do not have zero means if galaxies have locally preferred orientations on the sky. Such violation of statistical isotropy may result from extrinsic effects such as lensing shear, which biases $(\varepsilon_1,\,\varepsilon_2)$ but leaves $(q,\,u)$ unaffected~\citep{DyerShaver1992PolRotationByLens, Dai2014CMBPolRot}, or intrinsic shape alignment\footnote{Despite the standard nomenclature, intrinsic shape alignment is considered an {\it extrinsic} effect in our context. Unlike polarization-shape alignment, intrinsic shape alignment cannot be meaningfully quantified at the level of a single galaxy.}, which biases both $(q,\,u)$ and $(\varepsilon_1,\,\varepsilon_2)$. For a quantifiable model, we write
\begin{align}
\label{eq:pol_bias_IA}
    & \VEV{q} = - b_p\,\Gamma_1, \qquad \VEV{u} = - b_p\,\Gamma_2, \\
\label{eq:ellip_bias_IA}
    & \VEV{\varepsilon_1}  = \gamma_1 + \Gamma_1, \qquad \VEV{\varepsilon_2}  = \gamma_2 + \Gamma_2,
\end{align}
where the spin-2 quantities $(\gamma_1,\,\gamma_2)$ are the two components of the lensing shear, and the other spin-2 quantities $(\Gamma_1,\,\Gamma_2)$ are introduced to measure the ellipticity of the average intrinsic shape. 

\refEq{pol_bias_IA} and \refEq{ellip_bias_IA} indicate that intrinsic alignment results in nonzero means of the integral polarization $(q,\,u)$ along with nonzero means of the ellipticity $(\varepsilon_1,\,\varepsilon_2)$. The mean values are related through a bias coefficient $b_p$. Since this is a consequence of polarization-shape alignment, $b_p$ is related to $\delta$. If $\delta=\pi/2$, intrinsic alignment cannot affect the polarization statistics and we must have $b_p=0$. In the opposite limit, for perfect polarization-shape alignment with $\delta=0$, $b_p$ is maximized. 

Here we motivate the following simple choice for the bias coefficient: 
\begin{align}
\label{eq:bp_model}
    b_p=\frac{\sigma_p}{\sigma_\varepsilon}\,\cos\delta.
\end{align}
To understand this choice, let us consider a galaxy sample with constant, nonzero intrinsic shape alignment. The joint Gaussian PDF \refEq{fullPDF_noIA} for shape and polarization needs to be modified to allow the bias caused by intrinsic alignment. Inserting \refEq{bp_model} and performing straightforward algebra, we find
\begin{align}
\label{eq:fullPDF_withIA}
    & P(q,\,u,\,\varepsilon_1,\,\varepsilon_2|\Gamma_1,\,\Gamma_2)\,\rmd q\,\rmd u\,\rmd\varepsilon_1\,\rmd\varepsilon_2 = \nonumber\\
    & \frac{4}{(2\pi)^2\,\sigma^2_p\,\sigma^2_\varepsilon\,\sin^2\delta}\,\exp\left(-\frac{\Gamma^2_1+\Gamma^2_2}{2\,\sigma^2_\varepsilon}\right)\,\exp\left[- \frac{\varepsilon}{\sigma^2_\varepsilon}\,\left(\Gamma_1\,\cos 2\theta_e + \Gamma_2\,\sin 2\theta_e\right)\right]\nonumber\\
    & \times \exp\left[-\frac{\varepsilon^2}{2\,\sigma^2_\varepsilon\,\sin^2\delta} - \frac{p^2}{2\,\sigma^2_p\,\sin^2\delta} + \frac{\varepsilon\,p\,\cos\delta}{\sigma_p\,\sigma_\varepsilon\,\sin^2\delta}\,\cos2\delta\theta\right]\,\,p\,\rmd p\,\varepsilon\,\rmd\varepsilon\,\rmd\delta\theta\,\rmd\theta_e.
\end{align}
As expected, this modified distribution encodes preferred directions for $\theta_e$, which depend on $(\Gamma_1,\,\Gamma_2)$. 

We note a crucial property of \refEq{fullPDF_withIA}: for fixed polarization degree $p$ and ellipticity magnitude $\varepsilon$, the misalignment angle $\delta\theta$ has the same distribution as in \refEq{fullPDF_noIA} without intrinsic alignment, for the $(\Gamma_1,\,\Gamma_2)$-dependent factors in \refEq{fullPDF_withIA} do not involve $\delta\theta$. Physically, if the polarization-shape misalignment angle $\delta\theta$ only depends on the internal structure of the galaxy but not on its surrounding environment or any external influence, then intrinsic alignment is not supposed to alter the distribution of $\delta\theta$. This is indeed the case with \refEq{bp_model}. This nice mathematical property therefore justifies \refEq{bp_model} as the choice for the bias coefficient. In fact, directly calculating $P(q,\,u,\,\varepsilon_1,\,\varepsilon_2|\Gamma_1,\,\Gamma_2)\,\rmd q\,\rmd u\,\rmd\varepsilon_1\,\rmd\varepsilon_2$ for a general $b_p$ and transforming the variables to $(p,\,\varepsilon,\,\delta\theta,\,\theta_e)$ leads to the conclusion that \refEq{bp_model} is the {\it unique} choice for $b_p$ such that the distribution of $\delta\theta$ is not modified by nonzero $(\Gamma_1,\,\Gamma_2)$, as this choice renders terms in the exponent involving both $(\Gamma_1,\,\Gamma_2)$ and $\delta\theta$ vanishing. For the remainder of this paper, we shall use \refEq{bp_model}. 

\section{Likelihood} 
\label{sec:likelihood}

Building off the quantitative framework developed in previous sections, we now construct a likelihood model for the shape and polarization observables. Provided that the extrinsic effects are small perturbations, unbiased optimal quadratic estimators~\citep{Hu_2002, Maniyar_2021} can be derived from the principle of likelihood maximization~\citep{Hirata_2003weaklensing, Hirata_2003, Yin:2021kmx}, following an algebraic procedure familiar to the literature of CMB weak lensing. Below we start with the simple case in which statistics of galaxy shape and polarization are not altered by external effects.

We have assumed that $(\varepsilon_1,\,\varepsilon_2)$ and $(q,\,u)$ are correlated Gaussian random variables. This is different from the assumption made in \citep{yin2025new} that the products of $(\varepsilon_1,\,\varepsilon_2)$ and $(q,\,u)$ have Gaussian distributions. Ultimately, this assumption must be tested against real galaxy data. For now, this allows us to construct a simple likelihood function with analytic tractability.

Without extrinsic effects, the log likelihood function describing a single galaxy, up to an unimportant constant, is
\begin{align}
    \ln\mathcal{L}(q,\,u,\,\varepsilon_1,\,\varepsilon_2) = - \frac12\,\left(\frac{q^2 + u^2}{\sigma^2_p\,\sin^2\delta} + \frac{\varepsilon^2_1 + \varepsilon^2_2}{\sigma^2_\varepsilon\,\sin^2\delta} + 2\,\frac{q\,\varepsilon_1 + u\,\varepsilon_2}{\sigma_p\,\sigma_\varepsilon}\,\frac{\cos\delta}{\sin^2\delta} \right) - \ln\left(\sigma^2_p\,\sigma^2_\varepsilon\,\sin^2\delta\right).
\end{align}
This can be cast in the compact form
\begin{align}
    \ln\mathcal{L}(\bfx) = - \frac12\,\bfx^T\,\bfC^{-1}_0\,\bfx - \frac12\,\ln\left|\bfC_0\right|,
\end{align}
where we define a column vector $\bfx=[q,\,u,\,\varepsilon_1,\,\varepsilon_2]^T$, and the corresponding four-by-four covariance matrix:
\begin{align}
\label{eq:covC0}
    \bfC_0 = \begin{bmatrix}
        \sigma^2_p & 0 & - \sigma_p\,\sigma_\varepsilon\,\cos\delta & 0 \\
        0 & \sigma^2_p & 0 & - \sigma_p\,\sigma_\varepsilon\,\cos\delta \\
        - \sigma_p\,\sigma_\varepsilon\,\cos\delta & 0 & \sigma^2_\varepsilon & 0 \\
        0 & - \sigma_p\,\sigma_\varepsilon\,\cos\delta & 0 & \sigma^2_\varepsilon \\
    \end{bmatrix}.
\end{align}
The covariance matrix has a determinant $|\bfC_0|=\sigma^4_p\,\sigma^4_\varepsilon\,\sin^4\delta$. Its inverse is given by
\begin{align}
\label{eq:covC0_inv}
    \bfC^{-1}_0 = \begin{bmatrix}
        \frac{1}{\sigma^2_p\,\sin^2\delta} & 0 & \frac{\cos\delta}{\sigma_p\,\sigma_\varepsilon\,\sin^2\delta} & 0 \\
        0 & \frac{1}{\sigma^2_p\,\sin^2\delta} & 0 & \frac{\cos\delta}{\sigma_p\,\sigma_\varepsilon\,\sin^2\delta} \\
        \frac{\cos\delta}{\sigma_p\,\sigma_\varepsilon\,\sin^2\delta} & 0 & \frac{1}{\sigma^2_\varepsilon\,\sin^2\delta} & 0 \\
        0 & \frac{\cos\delta}{\sigma_p\,\sigma_\varepsilon\,\sin^2\delta} & 0 & \frac{1}{\sigma^2_\varepsilon\,\sin^2\delta} \\
    \end{bmatrix}.
\end{align}

Next, we consider how the single-galaxy likelihood is modified by extrinsic effects, which cause the observed polarization and ellipticity $\widetilde{\bfx} = [\widetilde q,\,\widetilde u,\,\widetilde \varepsilon_1,\,\widetilde \varepsilon_2]^T$ to differ from the intrinsic values $\bfx=[q,\,u,\,\varepsilon_1,\,\varepsilon_2]^T$. We model the simultaneous action of three distinct extrinsic effects:

\begin{enumerate}

\item {\bf Weak lensing shear} $(\gamma_1,\,\gamma_2)$ biases the distribution of ellipticity at the leading order as in \refEq{ellip_bias_IA}, but does not alter the integral polarization.

\item {\bf intrinsic shape alignment} $(\Gamma_1,\,\Gamma_2)$ biases the distributions of both ellipticity \refEq{ellip_bias_IA} and polarization \refEq{pol_bias_IA}.

\item Line-of-sight {\bf rotation of polarization direction} $\alpha$ mixes up the polarization Stokes variables $(q,\,u)$,
\begin{align}
\label{eq:alpha_effect}
    \widetilde q + i\,\widetilde u = (q + i\,u)\,e^{-2\,i\,\alpha},
\end{align}
but leaves ellipticity $(\varepsilon_1,\,\varepsilon_2)$ unchanged. Polarization rotation may be induced by new physics such as cosmic birefringence, come from any uncorrected Faraday rotation~\citep{Whittaker2017measuring}, or result from instrumental polarization miscalibration. 

\end{enumerate}

Expressing galaxy intrinsic shape and polarization in terms of the observed values, we derive the likelihood model that simultaneously captures all three extrinsic effects:
\begin{align}
\label{eq:lnL_all_three_effects}
    & \ln\mathcal{L}(\tilde q,\,\tilde u,\,\tilde \varepsilon_1,\,\tilde \varepsilon_2|\gamma_1,\,\gamma_2,\,\alpha, \Gamma_1, \Gamma_2) = \nonumber\\
    & - \frac{(\tilde q\,\cos2\alpha - \tilde u\,\sin2\alpha + b_p\,\Gamma_1)^2 + (\tilde u\,\cos2\alpha + \tilde q\,\sin2\alpha + b_p\,\Gamma_2)^2}{2\,\sigma^2_p\,\sin^2\delta} \nonumber\\
    & - \frac{(\tilde \varepsilon_1-\gamma_1 - \Gamma_1)^2 + (\tilde \varepsilon_2 - \gamma_2 - \Gamma_2)^2}{2\,\sigma^2_\varepsilon\,\sin^2\delta} \nonumber\\
    & - \frac{1}{\sigma_p\,\sigma_\varepsilon}\,\frac{\cos\delta}{\sin^2\delta}\,\left[ \left( \tilde q\,\cos 2\alpha - \tilde u\,\sin 2\alpha + b_p\,\Gamma_1 \right)\,(\tilde \varepsilon_1-\gamma_1 - \Gamma_1) \right.\nonumber\\
    & \left. + \left(\tilde u\,\cos 2\alpha + \tilde q\,\sin 2\alpha + b_p\,\Gamma_2 \right)\,(\tilde \varepsilon_2-\gamma_2 - \Gamma_2) \right].
\end{align}
It is clear from the likelihood that the effects of shear, polarization rotation, and intrinsic alignment can be distinguished from one another.

Finally, for a sample of galaxies with uncorrelated intrinsic properties (i.e. ellipticity and polarization), the total log likelihood function is trivially the sum of log likelihood functions for individual galaxies. Galaxies close to each other on the sky are subject to identical or highly similar extrinsic effects. Combining likelihood information from many galaxies will therefore enable measurements of these extrinsic effects.

\section{New estimators for shear, intrinsic alignment, and rotation} 
\label{sec:optest}

The analytic likelihood models presented in \refsec{likelihood} allows us to derive unbiased estimators of the extrinsic effects, which we turn to in this section.

We start by considering a simpler situation where intrinsic alignment can be safely neglected. In this case, only the two shear components plus the polarization rotation, i.e. $\hat\bftheta_3=[\hat\gamma_1,\,\hat\gamma_2,\,\hat\alpha]^T$, need to be simultaneously estimated. We need to set $\Gamma_1=\Gamma_2=0$ in the likelihood function \refEq{lnL_all_three_effects}. Following an analytic procedure similar to what was done in \cite{Yin:2021kmx, yin2025new} and outlined in \refapp{est_from_likelihood}, we derive the following unbiased, minimal-variance quadratic estimators\footnote{We refer to these as ``quadratic'' estimators, even though the estimators for shear and for intrinsic alignment are clearly linear in shape and polarization observables. Strictly speaking, the algebraic procedure in \refapp{est_from_likelihood} guarantees that the estimators are up to the quadratic order in shape and polarization observables.}
\begin{align}
\label{eq:hatbftheta3}
    \hat\bftheta_3 = \begin{bmatrix}
    \hat\gamma_1 \\
    \hat\gamma_2 \\
    \hat\alpha
    \end{bmatrix}
    = \begin{bmatrix}
        \widetilde \varepsilon_1 + \widetilde q\,\frac{\sigma_\varepsilon}{\sigma_p}\,\cos\delta \\
        \widetilde \varepsilon_2 + \widetilde u\,\frac{\sigma_\varepsilon}{\sigma_p}\,\cos\delta \\
        \left(\widetilde\varepsilon_1\,\widetilde u - \widetilde \varepsilon_2\,\widetilde q\right)/\left(4\,\sigma_p\,\sigma_\varepsilon\,\cos\delta\right) \\
    \end{bmatrix}.
\end{align}
Here, the shear components are estimated from the observed ellipticity components as is done traditionally in weak lensing, but the polarization observables are linearly combined as control variates that reduce variance (see \citep{2025arXiv251007375K} for an application of control variates in cosmology), even though shear does not change the polarization Stokes variables at all. This is enabled exactly by the correlation between ellipticity and polarization. On the other hand, the estimator for $\alpha$ is quadratic, since the action of polarization rotation is multiplicative.

The estimators in \refEq{hatbftheta3} have a covariance under the null hypothesis, i.e. in the absence of shear and polarization rotation,
\begin{align}
\label{eq:VEV_hatbftheta3_hatbftheta3}
    \VEV{\hat\bfthe_3\,\hat\bfthe^T_3} = \begin{bmatrix}
    \sigma^2_\varepsilon\,\sin^2\delta & 0 & 0 \\
    0 & \sigma^2_\varepsilon\,\sin^2\delta & 0 \\
     0 & 0 & \tan^2\delta/8 \\
    \end{bmatrix}.
\end{align}
This shows that the usual galaxy shape noise $\sigma_\varepsilon \sim 0.3$ is mitigated by a factor $\sin\delta < 1$.

Next, we consider the more general situation of non-negligible intrinsic alignment, using the likelihood \refEq{lnL_all_three_effects} with the full $(\Gamma_1,\,\Gamma_2)$ dependence. We need to simultaneously estimate five quantities, including two shear components, the angle of polarization rotation, and two components of intrinsic alignment. To this end, we form a vector $\hat\bftheta = [\hat\gamma_1,\,\hat\gamma_2,\,\hat\alpha,\,\hat\Gamma_1,\,\hat\Gamma_2]^T$. For the choice \refEq{bp_model}, we derive unbiased, minimal-variance estimators following the procedure in \refapp{est_from_likelihood}:
\begin{align}
\label{eq:hat_theta}
    \hat\bftheta = \begin{bmatrix}
    \hat\gamma_1 \\
    \hat\gamma_2 \\
    \hat\alpha \\
    \hat\Gamma_1 \\
    \hat\Gamma_2 \\
    \end{bmatrix}
    = \begin{bmatrix}
        \widetilde \varepsilon_1 + \widetilde q/b_p \\
        \widetilde \varepsilon_2 + \widetilde u/b_p \\
        \left(\widetilde\varepsilon_1\,\widetilde u - \widetilde \varepsilon_2\,\widetilde q\right)/\left(4\,\sigma_p\,\sigma_\varepsilon\,\cos\delta\right) \\
        - \widetilde q/b_p \\
        - \widetilde u/b_p \\
    \end{bmatrix}.
\end{align}

The shear estimators still linearly incorporate the polarization Stokes variables as control variates, in a way that is however different from \refEq{hatbftheta3}. The new estimators for intrinsic alignment linearly involve the polarization Stokes variables only. We also note that the estimators for shear and intrinsic alignment become singular if $b_p=0$, reflecting an exact degeneracy between the two if polarization information does not tell them apart.

It is useful to know the covariance of \refEq{hat_theta} under the null hypothesis, i.e. in the absence of all three extrinsic effects. For the choice \refEq{bp_model}, calculation yields
\begin{align}
\label{eq:VEV_hatheta_hattheta}
\VEV{\hat\bfthe\,\hat\bfthe^T} = \begin{bmatrix}
    \sigma^2_\varepsilon\,\tan^2\delta & 0 & 0 & - \sigma^2_\varepsilon\,\tan^2\delta & 0 \\
    0 & \sigma^2_\varepsilon\,\tan^2\delta & 0 & 0 & - \sigma^2_\varepsilon\,\tan^2\delta \\
    0 & 0 & \tan^2\delta/8 & 0 & 0 \\
    - \sigma^2_\varepsilon\,\tan^2\delta & 0 & 0 & \sigma^2_\varepsilon\,\sec^2\delta & 0 \\
    0 & - \sigma^2_\varepsilon\,\tan^2\delta & 0 & 0 & \sigma^2_\varepsilon\,\sec^2\delta \\
    \end{bmatrix}.
\end{align}
Estimations for shear and intrinsic alignment are correlated, while they are uncorrelated with estimating the angle of polarization rotation. Thus, our model predicts that the usual shape noise in estimating shear is mitigated by a factor $\tan\delta$ (which is less than unity if $\delta<\pi/4$). This amount of reduction is degraded compared to $\sin\delta$ in the case of neglecting intrinsic alignment, since there is a price to pay in order to tell shear and intrinsic alignment apart.

Ref.~\cite{Zhou2025TNG50polshape} sorted simulated star-forming galaxies into inclination bins and found $\delta=15^\circ$--$30^\circ$ for the top half of the star-forming galaxies that exhibit the best polarization-shape alignment. This corresponds to $\tan\delta = 0.27$--$0.58$ and $\sin\delta = 0.26$--$0.50$. Thus, allowing for nonzero intrinsic alignment effect safeguards against possible biases in shear estimation, at the cost of only modest degradation in the effective shape noise. 

\section{Measurement errors} 
\label{sec:noise}

In practice, ellipticity $(\varepsilon_1,\,\varepsilon_2)$ and polarization $(q,\,u)$ are not measured with infinite precision. To account for measurement errors, we replace the variances $\sigma^2_p$ and $\sigma^2_\varepsilon$ with tilded quantities, $\widetilde{\sigma}^2_p$ and $\widetilde{\sigma}^2_\varepsilon$. The latter are formed by adding measurement errors to the former in quadrature,
\begin{align}
\label{eq:wtilde_sig2_p}
    \widetilde\sigma^2_p & = \sigma^2_p + n^2_p, \\
\label{eq:wtilde_sig2_e}
    \widetilde\sigma^2_\varepsilon & = \sigma^2_\varepsilon + n^2_\varepsilon,
\end{align}
with the additional assumption that shape and polarization measurements have uncorrelated errors. We then denote a tilded correlation cosine
\begin{align}
    \cos\widetilde{\delta} = \frac{\sigma_p\,\sigma_\varepsilon}{\widetilde\sigma_p\,\widetilde\sigma_\varepsilon}\,\cos\delta < \cos\delta.
\end{align}
Since measurement errors inflate the variances, we must have $\widetilde\delta > \delta$. It can be seen that measurement errors modify the covariance matrix describing the shape and polarization observables, but do not modify their biases, which only depend on the parameter $b_p$.

Repeating a calculation in which $\sigma_p$, $\sigma_\varepsilon$ and $\delta$ are formally replaced with the corresponding tilded quantities, we find that the optimized estimators are given by \refEq{hat_theta} with $\cos\delta$ replaced with $\cos\widetilde\delta$, but the nonzero covariances under the null hypothesis are revised to
\begin{align}
\label{eq:covar_ga2}
    & \VEV{\hat\gamma^2_1} = \VEV{\hat\gamma^2_2} = \widetilde\sigma^2_\varepsilon + \frac{\widetilde\sigma^2_p}{b^2_p} - \frac{2\,\sigma_p\,\sigma_\varepsilon}{b_p}\,\cos\delta, \\
\label{eq:covar_alpha2}
    & \VEV{\hat\alpha^2} = \frac18\,\tan^2\widetilde\delta, \\
\label{eq:covar_Ga2}
    & \VEV{\hat\Gamma^2_1} = \VEV{\hat\Gamma^2_2} = \frac{\widetilde\sigma^2_p}{b^2_p}, \\
\label{eq:covar_gaGa}
    & \VEV{\hat\gamma_1\,\hat\Gamma_1} = \VEV{\hat\gamma_2\,\hat\Gamma_2} = - \frac{\widetilde\sigma^2_p}{b^2_p}\,\left(1 - \frac{b_p\,\sigma_p\,\sigma_\varepsilon}{\widetilde\sigma^2_p}\,\cos\delta \right).
\end{align}
These results are true for a general $b_p$, but in our model $b_p$ is still set by \refEq{bp_model}. As one might have anticipated, measurement errors inflate the variance of these estimators.

The estimators we construct for shear and for intrinsic alignment in \refEq{hat_theta} are not biased by polarization rotation at first order in external effects. However, polarization rotation, regardless of an astrophysical or instrumental origin, can contribute to their covariance at quadratic order even in the absence of shear or intrinsic alignment. This has not been included in the results \refeqs{covar_ga2}{covar_gaGa}. If polarization rotation is sizable, such second-order contributions are numerically important for setting up accurate reconstruction noise models, which are crucial for unbiased power spectrum estimation. Such a situation can easily arise, for example, if radio polarization miscalibration is on the order of a few degrees. This second-order contribution is analytically derived to be:
\begin{align}
\label{eq:covar_ga2_new}
    & \VEV{\hat\gamma^2_1} = \VEV{\hat\gamma^2_2} = \widetilde\sigma^2_\varepsilon + \frac{\widetilde\sigma^2_p}{b^2_p} - \frac{2\,\sigma_p\,\sigma_\varepsilon}{b_p}\,\cos\delta\,\left(1 - 2\,\alpha^2_{\rm rms}\right), \\
\label{eq:covar_alpha2_new}
    & \VEV{\hat\alpha^2} = \frac18\,\tan^2\widetilde\delta + 2\,\alpha^2_{\rm rms}, \\
\label{eq:covar_Ga2_new}
    & \VEV{\hat\Gamma^2_1} = \VEV{\hat\Gamma^2_2} = \frac{\widetilde\sigma^2_p}{b^2_p}, \\
\label{eq:covar_gaGa_new}
    & \VEV{\hat\gamma_1\,\hat\Gamma_1} = \VEV{\hat\gamma_2\,\hat\Gamma_2} = - \frac{\widetilde\sigma^2_p}{b^2_p}\,\left(1 - \frac{b_p\,\sigma_p\,\sigma_\varepsilon}{\widetilde\sigma^2_p}\,\cos\delta \right) + \frac{2\,\sigma_p\,\sigma_\varepsilon}{b_p}\,\cos\delta\,\alpha^2_{\rm rms}.
\end{align}
These are derived assuming that for the galaxy sample the polarization rotation angle $\alpha$ is a zero-mean Gaussian random quantity with a standard deviation $\alpha_{\rm rms}$. However, it should not be misinterpreted here that the polarization rotation necessarily has a white-noise angular power spectrum for these corrections to apply. Both physical or instrumental effects may imprint polarization rotations correlated across the sky. In principle, $\alpha_{\rm rms}$ can be treated as a free parameter informed by data. In this work, we will simply assume that $\alpha_{\rm rms}$ is known. Analytic results \refeqs{covar_ga2_new}{covar_gaGa_new} will be used in analyzing mock galaxy data in \refsec{mocks}.

\section{Mock estimation of angular power spectra} 
\label{sec:mocks}

\begin{figure*}[ht]
\includegraphics[width=\linewidth,scale=1.00]{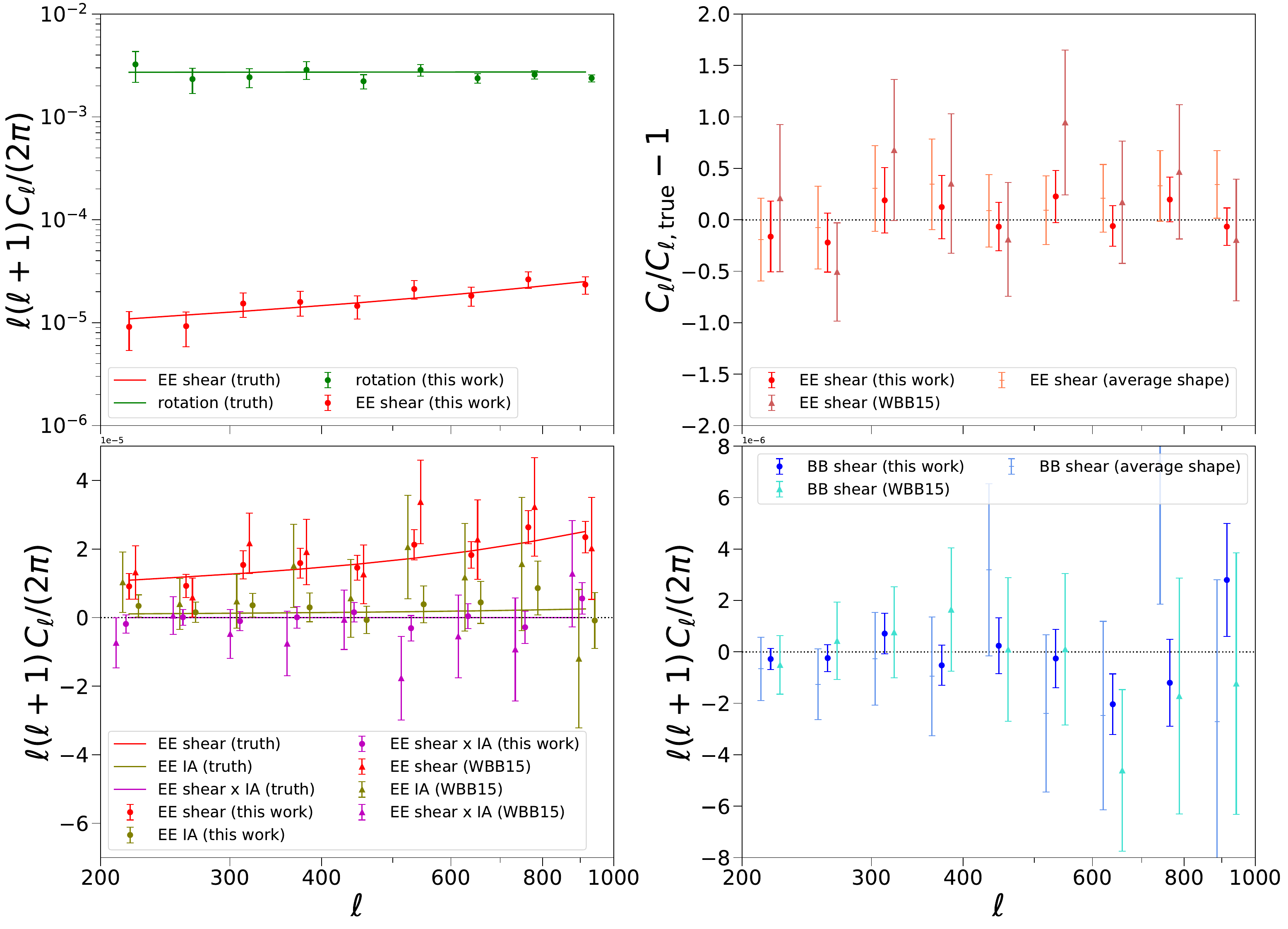}
\caption{Mock measurements of binned angular power spectra performed in a sky patch with $L=10\,{\rm deg}$, $n_g=10\,{\rm arcmin}^{-2}$ and $\delta=0.5$. For intrinsic alignment, we set $r_1=0.1$ and $r_2=0$. Lensing shear signals are injected assuming a fixed source redshift $z_s=1.0$. {\bf Upper left:} Reconstructed angular power spectra for E-mode shear and polarization rotation angle using the estimators developed in this work. {\bf Lower left:} Reconstructed angular auto power spectra for shear E modes and for intrinsic alignment E modes, as well as their angular cross power spectrum. {\bf Upper right:} Fractional reconstruction errors for the angular power spectrum of the shear E modes. {\bf Lower right:}  Reconstruction of the angular power spectrum of the shear B modes.}
\label{fig:Cls_mock_fiducial}
\end{figure*}

\begin{figure*}[h]
\includegraphics[width=\linewidth,scale=1.00]{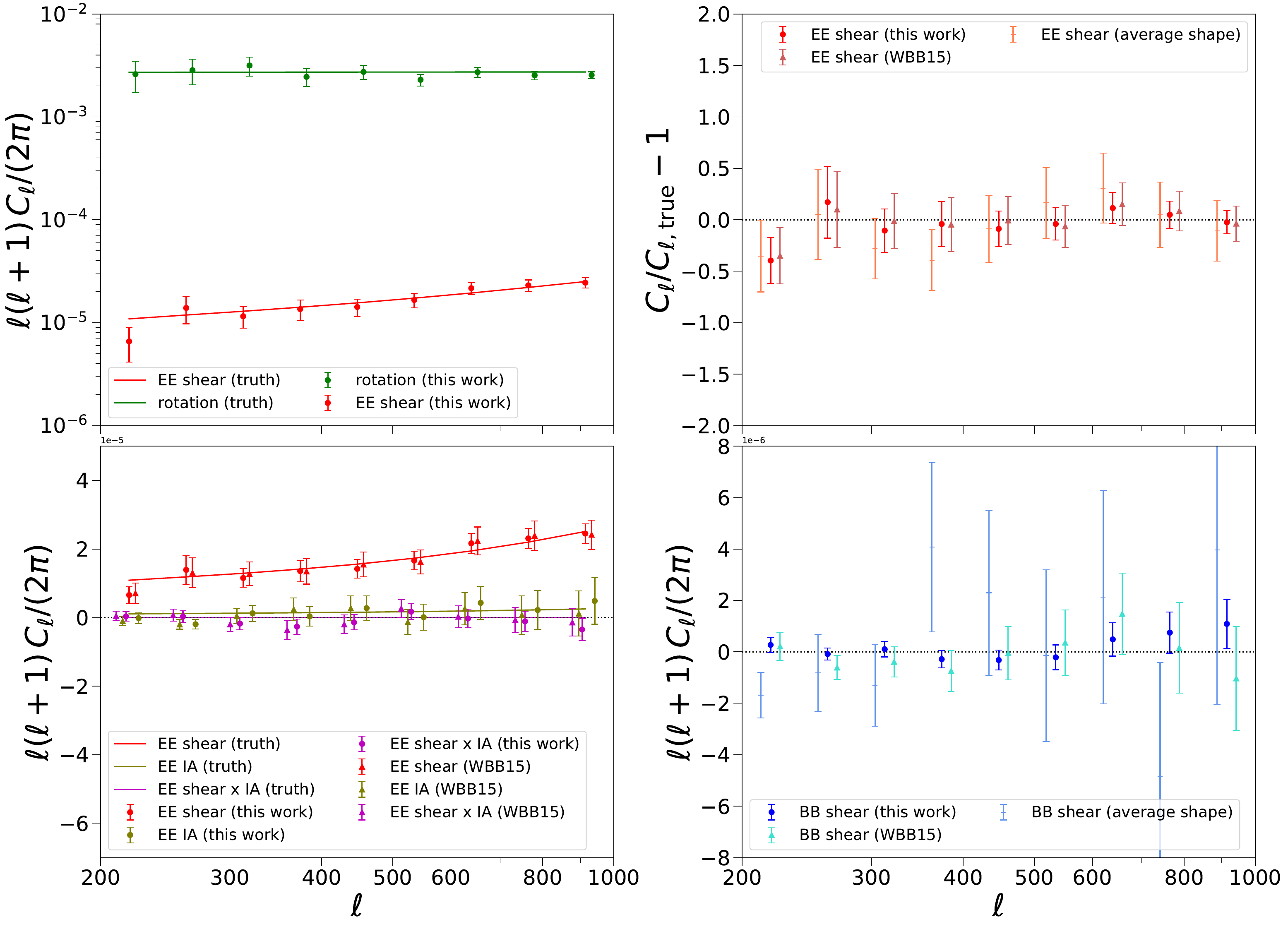}
\caption{Same as in \reffig{Cls_mock_fiducial} but assuming $\delta=0.1$. For the WBB15 estimators, uncertainties in angular power spectrum estimation are improved compared to simple shape averages.}
\label{fig:Cls_mock_fiducial_delta0d1}
\end{figure*}

To validate our new estimators and compare them to other estimators, below we generate mock measurements of galaxy shape and polarization. For a proof of concept, we will not attempt to model real-world observational systematics except for including Gaussian measurement errors in ellipticity and polarization. We do not mean to address the question whether the Gaussian model developed in \refsec{theory} is an adequate description of real galaxies. Rather, the tests will be presented merely to verify our analytic estimators. Throughout, we will adopt the Planck 2018 cosmological parameters~\citep{Aghanim2020Planck2018cosmology}.

The tests are carried out under the flat-sky approximation. Within a square sky footprint of side length $L$, we randomly place galaxies according to a mean surface number density $n_g$. Regarding the number of usable galaxies, a wide, shallow SKA-like continuum survey will yield $n_g=0.1$--$1\,{\rm arcmin}^{-2}$, while a deep-drilling survey may reach up to $n_g=10\,{\rm arcmin}^{-2}$~\citep{yin2025new}. For the purpose of computing shear signals, all mock galaxies are nominally at a single redshift $z_s$ but have randomized sky locations, with spatial clustering neglected. 

The mock galaxies are assigned to sky pixels of identical sizes. Within each pixel, extrinsic effects are assumed to be uniform: shear $(\gamma_1,\,\gamma_2)$, intrinsic alignment $(\Gamma_1,\,\Gamma_2)$, and polarization rotation $\alpha$. Gaussian random realizations of these extrinsic effects are generated on the sky, with spatial correlations properly implemented according to their angular power spectra. 

Particularly for the spin-2 shear variables $(\gamma_1,\,\gamma_2)$ and those describing intrinsic alignment $(\Gamma_1,\,\Gamma_2)$, we decompose them into E and B modes~\citep{Stebbins1996WeakLensingOnCelestialSphere, Cooray2002SecondOrderWeakLensing}. In the linear regime on large angular scales, the B modes vanish for the physical weak lensing shear~\citep{Schneider2002SmallScaleShearBmodes}. While both E and B modes may exist for intrinsic alignment~\citep{Hirata2004IALensing}, we assume the latter are vanishing. We compute a realistic angular power spectrum $C^{\gamma_E \gamma_E}_\ell$ for the E modes of lensing shear using the \texttt{halofit} model~\citep{Takahashi2012RevisedHaloFit} of nonlinear matter power spectrum. 

Instead of adopting realistic physical models, we prescribe the angular power spectra for intrinsic alignment and polarization rotation for the purpose of validation. The power spectrum for the rotation angle $\alpha$, which we assume to be uncorrelated with shear or intrinsic alignment, is set to be scale-invariant,
\begin{align}
    C^{\alpha\alpha}_\ell = 2\pi\,\alpha^2_0/\ell(\ell+1),
\end{align}
where $\alpha_0$ is a normalizing constant rotation angle. It is not necessary to specify if this represents physical rotation or miscalibration, as this is only intended for validating the estimators. 

For the E modes of intrinsic alignment, we not only set an auto power spectrum, but also allow a cross power spectrum with the shear E modes to mimic intrinsic alignment-shear interference expected for a finite redshift distribution of the shape sample~\citep{Hirata2004IALensing}. We assume a toy model in which both are proportional to the auto power spectrum of shear E modes:
\begin{align}
    C^{\Gamma_E \Gamma_E}_\ell = r_1\,C^{\gamma_E \gamma_E}_\ell, \qquad C^{\gamma_E \Gamma_E}_\ell = r_2\,C^{\gamma_E \gamma_E}_\ell,
\end{align}
where $r_1$ and $r_2$ are constants of proportionality. 

For each galaxy, a random set $(\varepsilon_1,\,\varepsilon_2,\,q,\,u)$ is drawn according to the multi-variate Gaussian model of \refsec{theory} defined by the covariance structure \refEq{qu_var}, \refEq{ellip_var} and \refEq{qu_ellip_covar}. Measurement noises for shear and polarization are included in the covariance through \refEq{wtilde_sig2_p} and \refEq{wtilde_sig2_e}. The bias coefficient $b_p$ is set with \refEq{bp_model}. Shear and intrinsic alignment effects are then implemented as biases in the Gaussian statistics, \refEq{pol_bias_IA} and \refEq{ellip_bias_IA}. Finally, polarization rotation is injected according to \refEq{alpha_effect}.

Assuming all mock galaxies carry the same statistical weight, shear, intrinsic alignment and polarization rotation in each sky pixel can be estimated by evaluating the estimators \refEq{hat_theta} for all galaxies within that pixel and then find their average. This yields pixelated, noisy sky maps of shear, intrinsic alignment and polarization rotation. Noises in these maps arise both from measurement errors in shape and polarization and from statistical variance intrinsic to the estimators themselves (c.f. \refEq{VEV_hatheta_hattheta}). From these noisy maps, angular power spectra and their statistical uncertainties can be estimated. Angular power spectra are estimated by maximizing the likelihood of the noisy maps, and statistical uncertainties are estimated from Fisher information. \refapp{est_power_spectra} presents calculational details of this process. In the likelihood function, a model for the noise covariance is required, for which we will either apply analytic results if available, or directly measure it from Monte Carlo samples.

\begin{figure*}[ht]
\includegraphics[width=\linewidth,scale=1.00]{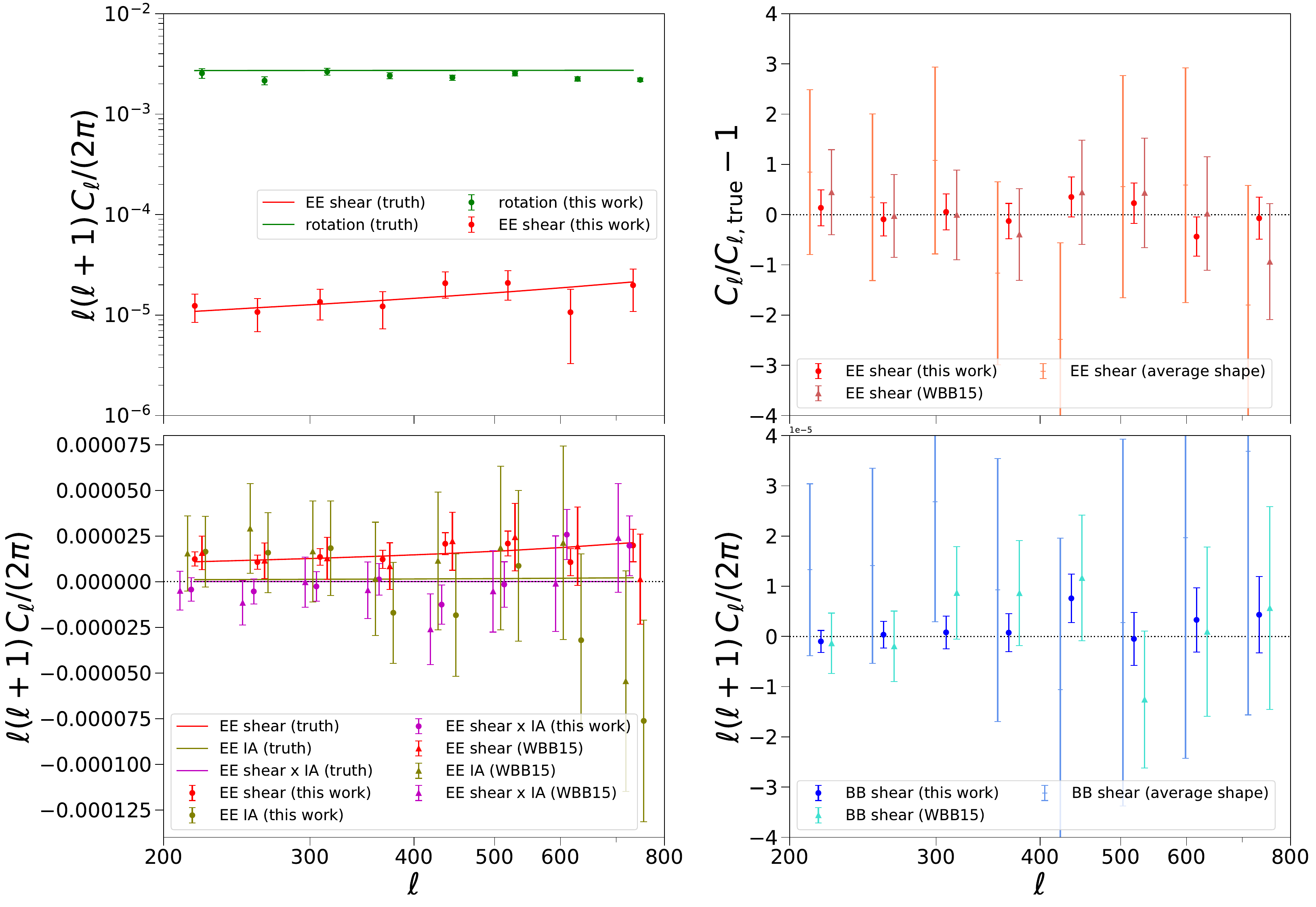}
\caption{Same as in \reffig{Cls_mock_fiducial} but for a hypothetical galaxy sample of high-quality polarization-shape alignment with $\delta=0.1$, from a mock survey of $L=30\,{\rm deg}$ and $n_g=0.3\,{\rm arcmin}^{-2}$. Thanks to the added radio polarization information, the new estimators in this work enable high precision in estimating the shear power spectra thanks to significantly mitigated shape noise.}
\label{fig:Cls_mock_delta0d1}
\end{figure*}

We compare three estimation methods:
\begin{enumerate}

\item \texttt{this work:} Unbiased, minimal-variance quadratic estimators for shear, polarization rotation and intrinsic alignment as we have developed from likelihood maximization.

\item \texttt{average shape:} The traditional shear estimators, which are simple averages of observed galaxy shapes and are biased by intrinsic alignment.

\item \texttt{WBB15:} The corrected shear estimators from \citep{Whittaker2015SeparateShearIA}, which are modified from the original shear estimators developed in \citep{BrownBattye2011weaklensingIA} and are devised, in the case of imperfect polarization-shape alignment, to remove a residual shear bias caused by intrinsic alignment. Corresponding estimators for intrinsic alignment are obtained as the difference between the simple shape average and the shear estimators.

\end{enumerate}

For the first two choices of estimators, analytic expressions for the corresponding noise covariance are available. For the WBB15 estimators, analytic noise covariance is unknown to us. In practice, the correct signal power spectra are not known a priori. Hence, accurate noise covariance can only be determined from iterations of Monte Carlo simulations, which was the approach in \citep{Whittaker2015SeparateShearIA}. To avoid expensive iterations of simulations, here we cheat by measuring noise covariance using simulated mock galaxies. While this practice would not be possible in real data analysis, doing so prevents us from unfairly introducing artificial biases in estimating the power spectra using the WBB15 method.

For the mock tests, we fix $\sigma_p=0.06$ and $\sigma_\varepsilon=0.26$. Since radio polarimetry SNRs will be the major limitation on the quality of measurement~\citep{yin2025new}, we include polarization measurement errors $n_p=0.02$ but neglect ellipticity measurement errors $n_\varepsilon=0$. 

\begin{figure*}[ht]
\includegraphics[width=\linewidth,scale=1.00]{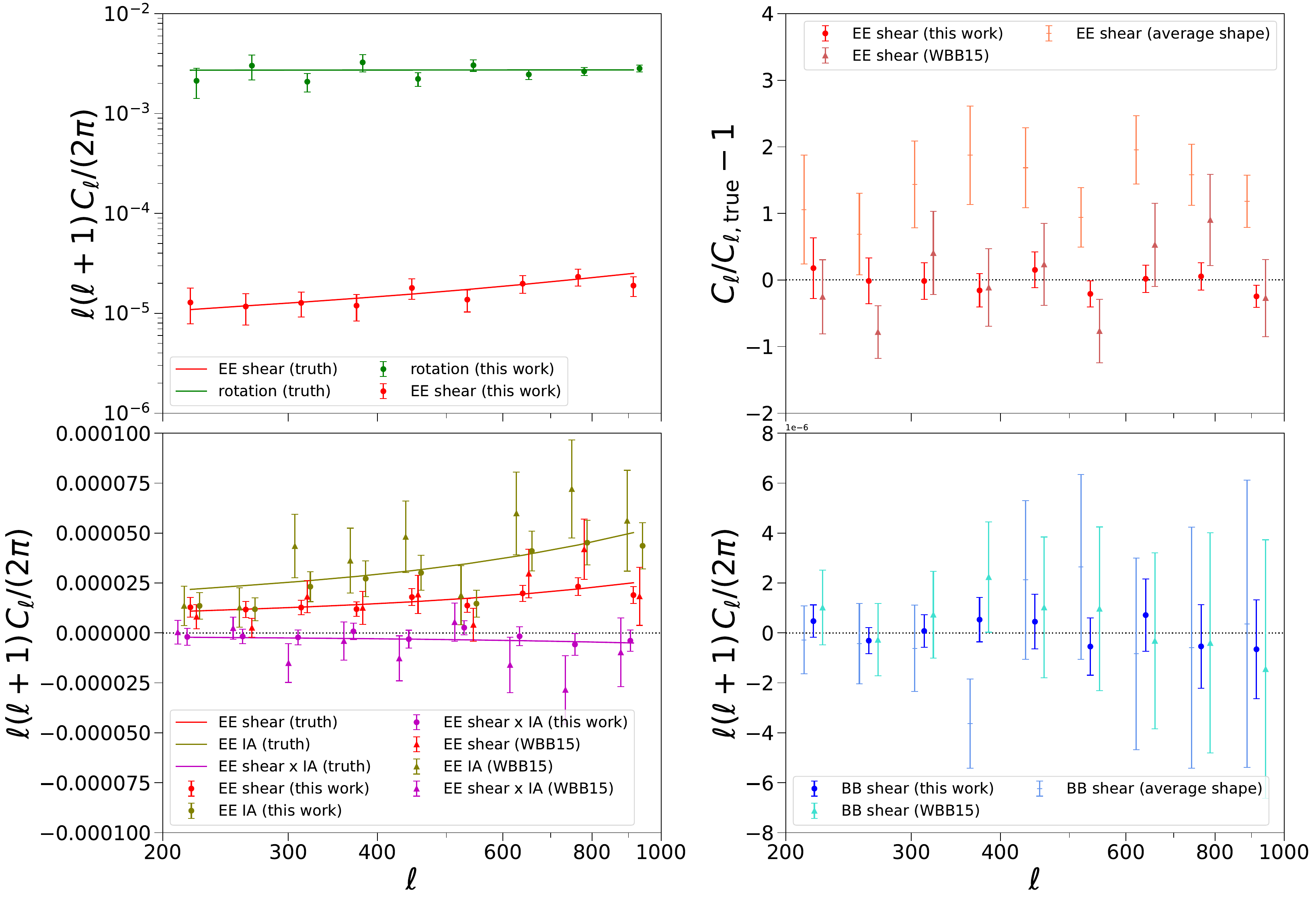}
\caption{Same as in \reffig{Cls_mock_fiducial}, but in a hypothetical situation of large intrinsic alignment with $r_1=2.0$ and $r_2=-0.2$. Simple average of galaxy shapes as the traditional shear estimators are biased by intrinsic alignment. The WBB15 shear estimators are unbiased, but the minimal-variance estimators developed in this work result in smaller noise in power spectrum estimation.}
\label{fig:Cls_mock_strongIA}
\end{figure*}

Mock measurements of power spectra are shown in \reffig{Cls_mock_fiducial}, for a deep survey of $100\,{\rm deg}^2$ with $n_g=10\,{\rm arcmin}^{-2}$. For polarization-shape alignment, we set a fiducial $\delta=0.5$, which roughly corresponds to what Ref.~\citep{Zhou2025TNG50polshape} found for the better half of the inclination bins at $z=0.7$--$1.0$ and observed at high frequencies ($\nu\geqslant 4.8\,$GHz). For intrinsic alignment, we set $r_1=0.1$ and $r_2=0$. Thus, the magnitude of the auto power spectrum for E-mode intrinsic alignment relative to that for the E-mode shear is similar to recent findings from cosmological hydrodynamic simulations~\citep{Ferlito2025IntrisicAlignmentMillenniumTNG}. Applying our new estimators with analytic reconstruction covariance, we achieve unbiased estimation for all angular power spectra. For shear and intrinsic alignment auto and cross power spectra, significantly smaller uncertainties are achieved than simple shape average or the WBB15 estimators. 

In \reffig{Cls_mock_fiducial}, the WBB15 estimators result in rather large errorbars compared to even the case of adopting simple shape average, contrary to examples presented in \citep{Whittaker2015SeparateShearIA}. This is due to the degraded quality of polarization-shape alignment $\delta=0.5$ assumed here than considered in \citep{Whittaker2015SeparateShearIA} (Gaussian random misalignment angle with a standard deviation $10^\circ$). Mock measurements performed for the same $L$ and $n_g$ but with $\delta=0.1$ lead to significantly decreased errorbars with the WBB15 estimators (but moderately larger than achievable with the new estimators in this work), as shown in \reffig{Cls_mock_fiducial_delta0d1}.

In \reffig{Cls_mock_delta0d1}, we perform similar mock measurements but instead consider a hypothetical galaxy sample with tight polarization-shape alignment $\delta=0.1$. For this, we simulate a wider survey of $900\,{\rm deg}^2$ with a low galaxy number density $n_g=0.3\,{\rm arcmin}^{-2}$. Unbiased power spectrum estimations with remarkably small uncertainties are again achieved with our new estimators.

\reffig{Cls_mock_strongIA} shows an alternative scenario that intrinsic alignment is much stronger, with $r_1=2$ and $r_2=-0.2$, for a survey over $100\,{\rm deg}^2$, with $n_g=10\,{\rm arcmin}^{-2}$ and $\delta=0.5$. As expected, the traditional shear estimators lead to a biased estimation of the E-mode shear power spectrum due to the confounding effect of intrinsic alignment. While the WBB15 estimators eliminate this bias, the uncertainties in power spectrum estimation are large. The results from our new estimators are free of biases and show the smallest uncertainties. 

\section{Conclusion} 
\label{sec:concl}

We have revisited the problem of measuring cosmic shear combining information about galaxy shape and integral radio polarization. Such galaxy samples are forthcoming through a synergy of optical imaging surveys and polarimetric radio continuum surveys. This problem is not only fundamental to reducing shape noise and separating lensing shear and intrinsic alignment, but is simultaneously relevant for detecting cosmic birefringence as a smoking gun of ultralight axions in the Universe.

We have introduced a model of galaxy shape and polarization observables as correlated Gaussian random quantities. In this model, polarization-shape alignment is captured by the covariance between shape and polarization, parameterized by a parameter $\delta$ which ranges from $\delta=0$ for perfect alignment to $\delta=\pi/2$ for no alignment at all. We have derived an analytic form \refEq{Delta_theta_pdf_model} that describes the non-Gaussian distribution of the polarization-shape misalignment angle. This was previously found to fit simulated galaxies well~\citep{Zhou2025TNG50polshape}, but has been elucidated for the first time here. A simple, analytically tractable likelihood function results from the correlated Gaussian model.

Following the principle of likelihood maximization, we have analytically derived a set of unbiased, minimal-variance estimators new to the literature (\refEq{hat_theta}), which allow simultaneous estimation of lensing shear, intrinsic shape alignment, and polarization rotation without degeneracy. These estimators take remarkably simple linear and quadratic forms in terms of the shape and polarization observables, and are accurate at linear order in the above three effects. The simple polynomial forms also guarantee numerical stability for few galaxies per sky element or for poor polarization-shape alignment (large $\delta$ values), which is not the case for other suggestions in the literature. Although we have emphasized on augmenting shear measurements, separately measuring polarization rotation and intrinsic shape alignment will be promising ways to uncover new physics.

For a useful analytic result, we have also derived the corresponding reconstruction noise covariance, \refeqs{covar_ga2}{covar_gaGa}, which facilitate unbiased measurements of angular power spectra from noisy map-level estimation using our estimators. Our analytic calculations have shown that adding polarization information reduces the shape noise by a factor $\tan\delta$ without any bias by intrinsic alignment. The same level of analytic simplicity and tractability is not seen in other independently proposed shear and intrinsic alignment estimators in the literature. We have demonstrated with toy galaxy mocks (\refsec{mocks}) how angular power spectra of shear, intrinsic alignment and rotation can be simultaneously and accurately measured using the new estimators, achieving minimized statistical uncertainties compared to results with alternative estimators.

The analytic results presented in this work therefore form an algorithmic foundation for future cosmology and/or fundamental physics applications based on the polarization-shape alignment effect.

\acknowledgments
The authors would like to thank Neal Dalal and Kendrick Smith for useful discussion. L.D. acknowledges research grant support from the Alfred P. Sloan Foundation (Award Number FG-2021-16495) and from the Office of Science, Office of High Energy Physics of the U.S. Department of Energy (Award Number DE-SC-0025293). Research at Perimeter Institute is supported in part by the Government of Canada through the Department of Innovation, Science and Economic Development and by the Province of Ontario through the Ministry of Colleges and Universities. R.Z. acknowledges the Berkeley Global Access (BGA) program, which enables a period of visitor scholarship that contributed to this work. 

\appendix

\section{Unbiased minimal-variance quadratic estimators from likelihood}
\label{app:est_from_likelihood}

Unbiased minimal-variance quadratic estimators have been the workhorse tools in many important inference problems in cosmology. Often referred to simply as the quadratic estimators~\citep{Hu_2002}, their construction and mathematical properties are well known in the literature of CMB weak lensing~\citep{Hu_2002, Maniyar_2021} and cosmic birefringence~\citep{Kamionkowski2009, Gluscevic2009, Yadav2009, Yin:2021kmx}. It has been known that these estimators can be derived from maximization of the likelihood in the perturbative regime, an approach demonstrated in Hirata \& Seljak~\citep{Hirata_2003}. In Ref.~\citep{Yin:2021kmx}, this approach is revisited for constructing quadratic estimators for polarization rotation in the CMB, which are conceptually identified with the so-called global-minimum-variance estimators~\citep{Maniyar_2021}. 

In \refsec{optest}, we have presented two sets of minimal-variance estimators following this same strategy. One is \refEq{hatbftheta3} for a situation where intrinsic alignment can be safely neglected and only shear and polarization rotation are estimated. The other is the more complete case \refEq{hat_theta} where intrinsic alignment is included as a physical effect and simultaneously estimated. 

For completeness, we outline in this Appendix the derivation of the more complete case, i.e. the estimators \refEq{hat_theta} based on the likelihood model \refEq{lnL_all_three_effects}. We will not repeat the algebra for the simpler case \refEq{hatbftheta3}, which can be derived following exactly the same algorithm but dropping the dependence on $(\Gamma_1,\,\Gamma_2)$ in \refEq{lnL_all_three_effects}. 

The first step is to perturbatively expand the log likelihood function of \refEq{lnL_all_three_effects} to quadratic orders in $\bftheta=[\gamma_1,\,\gamma_2,\,\alpha,\,\Gamma_1,\,\Gamma_2]^T$. This can be cast into the form
\begin{align}
    \ln\mathcal{L} \approx - \frac12\,\bftheta^T\,\bfM\,\bftheta + \bfb^T \bftheta + {\rm const.}.
\end{align}
Here, we shall perform calculations for a general bias parameter $b_p$. The 5-dimensional vector $\bfb$ vector has an explicit expression
\begin{align}
    \bfb = \begin{bmatrix}
       \frac{1}{\sigma^2_\varepsilon}\,\frac{1}{\sin^2\delta}\,\left(\widetilde\varepsilon_1 + \widetilde q\,\frac{\sigma_\varepsilon}{\sigma_p}\,\cos\delta\right) \\ 
       \frac{1}{\sigma^2_\varepsilon}\,\frac{1}{\sin^2\delta}\,\left(\widetilde\varepsilon_2 + \widetilde u\,\frac{\sigma_\varepsilon}{\sigma_p}\,\cos\delta\right) \\
       \frac{2}{\sigma_p\,\sigma_\varepsilon}\,\frac{\cos\delta}{\sin^2\delta}\,\left(\widetilde\varepsilon_1\,\widetilde u - \widetilde \varepsilon_2\,\widetilde q\right)\\
       \frac{1}{\sigma^2_\varepsilon}\,\frac{1}{\sin^2\delta}\,\left( \widetilde\varepsilon_1 - \frac{b_p}{\sigma^2_p}\,\widetilde q + \frac{\cos\delta}{\sigma_p\,\sigma_\varepsilon}\,\left(\widetilde q - b_p\,\widetilde\varepsilon_1 \right) \right) \\
       \frac{1}{\sigma^2_\varepsilon}\,\frac{1}{\sin^2\delta}\,\left( \widetilde\varepsilon_2 - \frac{b_p}{\sigma^2_p}\,\widetilde u + \frac{\cos\delta}{\sigma_p\,\sigma_\varepsilon}\,\left(\widetilde u - b_p\,\widetilde\varepsilon_2 \right) \right) \\
    \end{bmatrix}.
\end{align}
From direct calculation, it is found that the 5$\times$5 symmetric matrix $\bfM$ has the following nonzero matrix elements
\begin{align}
    & {\rm M}_{\gamma_1\gamma_1} = {\rm M}_{\gamma_2\gamma_2} = 1/\left(\sigma^2_\varepsilon\,\sin^2\delta\right), \\
    & {\rm M}_{\alpha\alpha} = - \frac{4}{\sigma_p\,\sigma_\varepsilon}\,\frac{\cos\delta}{\sin^2\delta}\, \left(\widetilde\varepsilon_1\,\widetilde q + \widetilde\varepsilon_2\,\widetilde u \right), \\
    & {\rm M}_{\Gamma_1\Gamma_1} = {\rm M}_{\Gamma_2\Gamma_2} = \frac{1}{\sigma^2_\varepsilon\,\sin^2\delta}\,\left(1 + \frac{b^2_p\,\sigma^2_\varepsilon}{\sigma^2_p} - \frac{2\,b_p\,\sigma_\varepsilon}{\sigma_p}\,\cos\delta\right), \\
    & {\rm M}_{\gamma_1\alpha} = \frac{2}{\sigma_p\,\sigma_\varepsilon}\,\frac{\cos\delta}{\sin^2\delta}\,\widetilde u, \qquad {\rm M}_{\gamma_2\alpha} = - \frac{2}{\sigma_p\,\sigma_\varepsilon}\,\frac{\cos\delta}{\sin^2\delta}\, \widetilde q, \\
    & {\rm M}_{\gamma_1\Gamma_1} = {\rm M}_{\gamma_2\Gamma_2} = \frac{1}{\sigma^2_\varepsilon\,\sin^2\delta}\,\left(1 - b_p\,\frac{\sigma_\varepsilon}{\sigma_p}\,\cos\delta\right), \\
    & {\rm M}_{\alpha\Gamma_1} = \frac{2\,(\cos\delta - b_p\,\frac{\sigma_\varepsilon}{\sigma_p})}{\sigma_p\,\sigma_\varepsilon\,\sin^2\delta}\,\widetilde u, \qquad {\rm M}_{\alpha\Gamma_2} = - \frac{2\,(\cos\delta - b_p\,\frac{\sigma_\varepsilon}{\sigma_p})}{\sigma_p\,\sigma_\varepsilon\,\sin^2\delta}\,\widetilde q,
\end{align}
which can depend on shape and polarization observables. In the perturbative regime, we must take the ensemble average of this matrix $\bfM$, evaluated under the null hypothesis, i.e. in the absence of lensing shear, intrinsic alignment and polarization rotation. The ensemble average is worked out to be
\begin{align}
    \overline \bfM = \begin{bmatrix}
        \frac{1}{\sigma^2_\varepsilon\,\sin^2\delta} & 0 & 0 & \frac{1-b_p\,\frac{\sigma_\varepsilon}{\sigma_p}\,\cos\delta}{\sigma^2_\varepsilon\,\sin^2\delta} & 0 \\
        0 & \frac{1}{\sigma^2_\varepsilon\,\sin^2\delta} & 0 & 0 & \frac{1-b_p\,\frac{\sigma_\varepsilon}{\sigma_p}\,\cos\delta}{\sigma^2_\varepsilon\,\sin^2\delta} \\
        0 & 0 & \frac{8\,\cos^2\delta}{\sin^2\delta} & 0 & 0 \\
        \frac{1-b_p\,\frac{\sigma_\varepsilon}{\sigma_p}\,\cos\delta}{\sigma^2_\varepsilon\,\sin^2\delta} & 0 & 0 & \frac{1 + \frac{b^2_p\,\sigma^2_\varepsilon}{\sigma^2_p} - \frac{2\,b_p\,\sigma_\varepsilon}{\sigma_p}\,\cos\delta}{\sigma^2_\varepsilon\,\sin^2\delta} & 0 \\ 
        0 & \frac{1-b_p\,\frac{\sigma_\varepsilon}{\sigma_p}\,\cos\delta}{\sigma^2_\varepsilon\,\sin^2\delta} & 0 & 0 & \frac{1 + \frac{b^2_p\,\sigma^2_\varepsilon}{\sigma^2_p} - \frac{2\,b_p\,\sigma_\varepsilon}{\sigma_p}\,\cos\delta}{\sigma^2_\varepsilon\,\sin^2\delta}
    \end{bmatrix}.
\end{align}
This matrix can be analytically inverted
\begin{align}
    \overline \bfM^{-1} = \begin{bmatrix}
    \frac{\sigma^2_p}{b_p^2} + \sigma^2_\varepsilon - \frac{2\,\sigma_p\,\sigma_\varepsilon}{b_p}\,\cos\delta & 0 & 0 & \frac{\sigma^2_p}{b_p^2}\,\left( \frac{b_p\,\sigma_\varepsilon}{\sigma_p}\,\cos\delta - 1 \right) & 0 \\
    0 & \frac{\sigma^2_p}{b_p^2} + \sigma^2_\varepsilon - \frac{2\,\sigma_p\,\sigma_\varepsilon}{b_p}\,\cos\delta & 0 & 0 & \frac{\sigma^2_p}{b_p^2}\,\left( \frac{b_p\,\sigma_\varepsilon}{\sigma_p}\,\cos\delta - 1 \right) \\
    0 & 0 & \frac{\tan^2\delta}{8} & 0 & 0 \\
     \frac{\sigma^2_p}{b_p^2}\,\left( \frac{b_p\,\sigma_\varepsilon}{\sigma_p}\,\cos\delta - 1 \right) & 0 & 0 & \frac{\sigma^2_p}{b_p^2} & 0 \\
     0 & \frac{\sigma^2_p}{b_p^2}\,\left( \frac{b_p\,\sigma_\varepsilon}{\sigma_p}\,\cos\delta - 1 \right) & 0 & 0 & \frac{\sigma^2_p}{b_p^2} \\ 
    \end{bmatrix}.
\end{align}
The log likelihood function is then approximated as
\begin{align}
\ln\mathcal{L} \approx - \frac12\,\bftheta^T\,\overline\bfM\,\bftheta + \bfb^T \bftheta + {\rm const.}.
\end{align}
This quadratic expression is maximized for
\begin{align}
\label{eq:bftheta_Minvb_AppA}
    \bftheta = \overline\bfM^{-1}\bfb.
\end{align}
Evaluating $\overline\bfM^{-1}\bfb$ reproduces the expressions \refEq{hat_theta}.

Under the null hypothesis, these estimators have covariance
\begin{align}
\left\langle\hat\bftheta\,\hat\bftheta^T\right\rangle = \overline\bfM^{-1}\left\langle\bfb\,\bfb^T\right\rangle\,\overline\bfM^{-1} = \overline\bfM^{-1}\,\overline\bfM\,\overline\bfM^{-1} = \overline\bfM^{-1}.
\end{align}
Evaluating this and making the choice \refEq{bp_model}, we reproduce \refEq{VEV_hatheta_hattheta}.

When intrinsic alignment effects are neglected, \refEq{hatbftheta3} and \refEq{VEV_hatbftheta3_hatbftheta3} give the quadratic estimators and their covariance under the null hypothesis, respectively. These can be derived following the same logic, but by setting $\Gamma_1=\Gamma_2=0$ in the likelihood function in the first place.

\section{Estimating power spectra for multiple Gaussian random quantities}
\label{app:est_power_spectra}

In this work, just like in many cosmology inference problems, we need to estimate the angular (auto- and cross-)power spectra of some zero-mean Gaussian random quantities on the sky. As a matter of practice, we shall estimate those in bins of the angular wave number $\ell$, with the approximation that in each $\ell$ bin both the signal and noise (co-)variances are the same for all Fourier modes on the sky. Thus, different Fourier modes provide independent estimates for the (co-)variance under the general assumption of statistical isotropy. In this Appendix, we collect analytic results on how the angular power spectra can be estimated from noisy measurement of the Gaussian random quantities.

It is useful to first review the result for a single Gaussian random quantity. Suppose that in a given $\ell$ bin there are $N$ statistically independent, real-valued multipole moments, which we denote as $d_i$ for $i=1,\,2,\,\cdots,N$. It is well known that the optimal estimation for the signal variance is given by $\widehat{\sigma^2} = \left(\sum^N_{i=1}\,d^2_i\right) - n^2$, where $n^2$ is the corresponding noise variance. This textbook result can be derived from maximization of the likelihood.

We now discuss the case of multiple Gaussian random quantities. In the context of this work, we need to estimate the angular power spectra involving five quantities $(\gamma_1,\,\gamma_2,\,\alpha,\,\Gamma_1,\,\Gamma_2)$, the estimation of which are correlated. In a given $\ell$ bin, we have one measurement of these five quantities from each Fourier mode on the sky, with a total of $N$ independent measurements. These measurements are represented by vectors $\bfd_i$ for $i=1,\,2,\cdots,N$. Let $\bfD$ be the signal covariance matrix, and $\bfN$ be the noise covariance matrix. Both are $5\times 5$ positive matrices. In principle, $\bfN$ would include both the variance of the estimation noise alone (i.e. the difference between the estimator and the true signal) and the covariance between the estimation noise and the true signal~\citep{Whittaker2015SeparateShearIA}. In our tests, however, the latter is found to be numerically negligible compared to the former.  

We first consider the case that the signal covariance $\bfD$ is diagonal. Provided that $\bfN$ can be computed from \refeqs{covar_ga2_new}{covar_gaGa_new}, what is the optimal estimate of $\bfD$ combining all $N$ measurements in the given $\ell$ bin? This question can be answered from maximization of the likelilhood. The log likelihood function is
\begin{align}
    \ln L = - \frac{N}{2}\,\ln{\rm det}\left(\bfN + \bfD \right) - \frac12\,\sum^N_{i=1}\,\bfd^T_i\left(\bfN + \bfD\right)^{-1} \bfd_i.
\end{align}
Let $p_\alpha$ be the $\alpha$-th diagonal element of $\bfD$, for $\alpha=1,\,2,\,\cdots,\,5$, which gives the signal variance of the $\alpha$-th Gaussian random quantity. Maximization of the likelihood requires
\begin{align}
    0 = \frac{\partial \ln L}{\partial p_\alpha} =  - \frac{N}{2}\,{\rm Tr}\left[\left(\bfN + \bfD \right)^{-1}\,\partial_\alpha\bfD \right] + \frac12\,\sum^N_{i=1}\,\bfd^T_i \left(\bfN + \bfD \right)^{-1}\,\partial_\alpha\bfD\,\left(\bfN + \bfD \right)^{-1} \bfd_i,
\end{align}
where we assume that $\bfN$ does not depend on the $p_\alpha$'s. Since $p_\alpha$'s are the diagonal elements of $\bfD$, this reduces to one equation for each index $\alpha$,
\begin{align}
\label{eq:AppB_MLE_eqn}
    \left(\bfN + \bfD \right)^{-1}_{\alpha\alpha} = \frac{1}{N}\,\sum^N_{i=1}\,\left[ \left(\bfN + \bfD \right)^{-1}\bfd_i\right]^2_\alpha, \qquad \alpha = 1,\,2,\,\cdots,\,5.
\end{align}
When $\bfN$ is in general non-diagonal, these questions form a coupled set for the $p_\alpha$'s, which in general need to be solved numerically. Covariance in this estimation can be quantified using the Fisher information matrix~\citep{Hirata_2003}, which is
\begin{align}
\label{eq:AppB_Fisher_mat}
    F_{\alpha\beta} = \left\langle-\partial_\alpha\partial_\beta\ln L\right\rangle = \frac{N}{2}\,{\rm Tr}\left[\left(\bfN + \bfD \right)^{-1}\,\partial_\beta\bfD\,\left(\bfN + \bfD \right)^{-1}\,\partial_\alpha\bfD\right] = \frac{N}{2}\,\left(\bfN + \bfD \right)^{-1}_{\alpha\beta}\,\left(\bfN + \bfD \right)^{-1}_{\beta\alpha}.
\end{align}
In the second line, we have simplified the matrix algebra under the assumption that $p_\alpha$'s are the diagonal elements of $\bfD$. In the numerical tests we present in \refsec{mocks}, \refEq{AppB_MLE_eqn} is used to calculate the angular (auto- and cross-)power spectra, and \refEq{AppB_Fisher_mat} is used to estimate the corresponding statistical uncertainties.

More generally, if we estimate some parameters $p_\alpha$'s which may enter both the signal covariance $\bfD$ and the noise covariance $\bfN$, then maximal likelihood estimation can be solved from the following set of coupled equations,
\begin{align}
    {\rm Tr}\left[\left(\bfN + \bfD\right)^{-1}\,\partial_\alpha\left(\bfN + \bfD \right)\right] = \frac{1}{N}\,\sum^N_{i=1}\,\bfd^T_i\,\left(\bfN + \bfD\right)^{-1}\,\partial_\alpha\left(\bfN + \bfD \right)\,\left(\bfN + \bfD\right)^{-1}\,\bfd_i,
\end{align}
one for each value of $\alpha$. The corresponding Fisher matrix has an expression
\begin{align}
    F_{\alpha\beta} = \frac{N}{2}\,{\rm Tr}\left[\left(\bfN + \bfD\right)^{-1}\,\partial_\alpha\left(\bfN + \bfD \right)\,\left(\bfN + \bfD\right)^{-1}\,\partial_\beta\left(\bfN + \bfD \right)\right].
\end{align}


\bibliographystyle{JHEP}
\bibliography{radio_pol}

\end{document}